\documentclass[sigconf]{acmart}

\usepackage{color}
\usepackage{array}
\usepackage{graphicx}
\usepackage{soul}
\usepackage{colortbl} 
\usepackage{multirow}
\usepackage{array}

\definecolor{lightyellow}{RGB}{245,231,178}
\sethlcolor{lightyellow}

\soulregister\cite7
\soulregister\ref7
\soulregister\autoref7
\soulregister\citeauthor7
\newcommand{\rhl}[1]{#1}
\newcommand{\rhlp}[1]{#1}

\AtBeginDocument{%
  \providecommand\BibTeX{{%
    \normalfont B\kern-0.5em{\scshape i\kern-0.25em b}\kern-0.8em\TeX}}}

\copyrightyear{2025}
\acmYear{2025}
\setcopyright{acmlicensed}\acmConference[CHI '25]{CHI Conference on Human Factors in Computing Systems}{April 26-May 1, 2025}{Yokohama, Japan}
\acmBooktitle{CHI Conference on Human Factors in Computing Systems (CHI '25), April 26-May 1, 2025, Yokohama, Japan}
\acmDOI{10.1145/3706598.3713219}
\acmISBN{979-8-4007-1394-1/25/04}

\begin{document}

\title[Understanding and Supporting Peer Review Using AI-reframed Positive Summary]{Understanding and Supporting Peer Review Using AI-reframed Positive Summary}


\author{Chi-Lan Yang}
\email{chilan.yang@iii.u-tokyo.ac.jp}
\orcid{0000-0003-0603-2807}
\affiliation{%
  \institution{The University of Tokyo, Graduate School of Interdisciplinary Information Studies}
  \city{Tokyo}
  \country{Japan}
}

\author{Alarith Uhde}
\email{pcs@alarithuhde.com}
\orcid{0000-0003-3877-5453}
\affiliation{%
  \institution{The University of Tokyo, Tokyo College}
  \city{Tokyo}
  \country{Japan}
}

\author{Naomi Yamashita}
\email{naomiy@acm.org}
\orcid{0000-0003-0643-6262}
\affiliation{%
  \institution{1: NTT Communication Science Laboratories; 2: Kyoto University, Social Informatics}
  \city{1: Keihanna; 2: Kyoto}
  \country{Japan}
}

\author{Hideaki Kuzuoka}
\email{kuzuoka@cyber.t.u-tokyo.ac.jp}
\orcid{0000-0003-1252-7814}
\affiliation{%
  \institution{The University of Tokyo, Graduate School of Information Science and Technology}
  \city{Tokyo}
  \country{Japan}
}

\renewcommand{\shortauthors}{Yang, et al.}

\begin{abstract}
While peer review enhances writing and research quality, harsh feedback can frustrate and demotivate authors. Hence, it is essential to explore how critiques should be delivered to motivate authors and enable them to keep iterating their work. In this study, we explored the impact of appending an automatically generated positive summary to the peer reviews of a writing task, alongside varying levels of overall evaluations (high vs. low), on authors’ feedback reception, revision outcomes, and motivation to revise. Through a 2x2 online experiment with 137 participants, we found that adding an AI-reframed positive summary to otherwise harsh feedback increased authors’ critique acceptance, whereas low overall evaluations of their work led to increased revision efforts. We discuss the implications of using AI in peer feedback, focusing on how AI-driven critiques can influence critique acceptance and support research communities in fostering productive and friendly peer feedback practices.
\end{abstract}




\maketitle

\section{Introduction}
Peer reviews are crucial for quality and validity in scientific knowledge production, as they involve domain experts evaluating their peers' work \cite{kelly2014peerreview}. 
However, this process is not without challenges, especially when it comes to delivering feedback. 
Unclear review guidelines, missed opportunities to follow review guidelines properly \cite{sun2024reviewflow_IUI24}, the anonymity of the peer review \cite{kaya2021anonymouspeerfeedback_2021, walsh2000open}, heavy workload, lacking support for novice reviewers \rhlp{\mbox{\cite{sun2024metawriter_CSCW24}}}, and unclear review norms within the community \cite{jansen2016CHIreview} can sometimes lead to non-constructive or harsh criticism, which may be perceived as unfair or overly negative by the recipients \cite{Nguyen2017fruitfulfeedback, jansen2016CHIreview}.
Such critical feedback can demotivate recipients (i.e., authors), potentially leading them to cease efforts to improve their work.

Despite the recognized importance of providing constructive feedback, the current practice in research communities places a significant cognitive burden on feedback providers (i.e., reviewers). 
They are expected not only to provide professional and high-quality reviews but also to offer critiques in a positive and constructive tone \cite{jansen2016CHIreview, kelly2014peerreview, pnas_reviewer_guidelines, chi2024_reviewing_guide, acm_tochi_reviewer_guidelines, cscw_mentor_program, vrst2023_reviewing, agu_review_criteria}.
This requirement demands additional effort, especially when the review workload is already heavy. 
This is where large language models (LLMs) can play a supportive role by assisting in reframing critical feedback in a more positive manner. 
Recent advancements in LLMs have made their application in the peer review process increasingly feasible \rhlp{\mbox{\cite{liang2024LLMfeedback_arxiv2024, sun2024metawriter_CSCW24}}}, particularly for tasks such as paraphrasing existing content\footnote{https://www.grammarly.com/paraphrasing-tool}. 
\rhl{When incorporating LLMs into the peer review process, it is crucial to effectively utilize LLMs while maintaining our originality and the autonomy that comes with being a human reviewer. 
Recent works on AI-assisted writing and AI-mediated communication indicated that people felt a diminished sense of control when they had minimal influence over the final outcomes of AI-assisted writing~\cite{draxler2024aighostwriter}.
Additionally, they experienced a lack of authenticity when composing messages with AI for communication purposes~\cite{fu2024text}.
Thus, instead of completely offloading the writing task to LLMs, we aim to investigate the possibility of using LLMs to ``reframe'' content based on what humans have written.} 
We developed and evaluated the concept of an AI-reframed summary that reframes critical feedback more positively and appends it to the original critique written by the reviewers to help authors engage more constructively with the review.

\rhl{
The main focus of this study was to see how AI-reframed positive summaries affected authors' perceptions and willingness to revise.
Meanwhile, we cannot ignore the presence of ``overall evaluation'' scores, which are prevalent in peer reviews.
In the current practice of peer review, authors receive critique with an ``overall evaluation'' of their work, which often includes scores or overall recommendations such as ``accept with minor revision,'' ``major revision,'' or ``reject'' \cite{chi2024_reviewing_guide, acm_tochi_reviewer_guidelines, siggraph2024_reviewer_instructions}.
The ``overall evaluation'' helps editors and reviewers make collective decisions, but related works have shown that receiving scores influenced authors motivation and task performance, even if scores were randomly assigned~\cite{zhou1998feedback}.
When we append an AI-reframed positive summary to critique that comes with an ``overall evaluation'', how these factors influence authors' further revision of the work is unclear.
Hence, this study aims to examine how these scores interact with AI-reframed positive summaries. 
}


Our study examined how AI-reframed positive summaries, combined with overall evaluations, influenced the way authors react to critiques.
\rhl{We define ``overall evaluation'' in this study as the perceived \textit{relative performance} of the authors receiving feedback for their work, compared to other unknown peers. 
The evaluation scores are not necessarily linked to the actual quality of their work, as different reviewers may have varying standards.}
We conducted a two-by-two controlled experiment with 137 participants, varying the presence of \textit{AI-reframed positive summaries} (with vs. without) and the \textit{overall evaluation ratings} (high vs. low) to explore how these factors influenced participants' critique reception differently. 
Our findings show that critiques accompanied by an AI-reframed positive summary and a high overall evaluation led to increased positive emotions among authors. 
Moreover, these AI-reframed positive summaries increased authors' perceived autonomy and competence, and improved their perceived fairness of the feedback and the reviewers.
Interestingly, while showing AI-reframed positive summaries did not significantly impact revision efforts, receiving a low overall evaluation did lead to significantly more revision efforts. 
Thematic analysis of the open-ended comments in the survey using the \textit{attributional theory of motivation} revealed that both \textit{intrapersonal} and \textit{interpersonal} factors authors perceived from the critique influenced their motivation to revise their work.

This work makes three key contributions: First, it advances the understanding of how LLMs can be integrated into the peer review process by identifying both the opportunities and challenges involved; second, it provides empirical evidence on the effects of AI-generated positive summaries of reviewers' critical feedback and overall evaluation scores on feedback reception and authors' subsequent revision efforts; third, it offers broader insights into the implications of using LLMs in peer review and online feedback systems while suggesting directions for future research to further optimize and expand their role in enhancing these processes.

\section{Background and Related Works}
\subsection{Supporting Peer Feedback}
\label{sec:tech_supported_peer_review}
Peer review serves critical functions for both individuals and the community \cite{huang2016conflictpeerreview_CSCW16}. 
From an individual perspective, receiving peer feedback allows authors to gain new knowledge and refine and improve their work. 
From a community perspective, peer review acts as a gatekeeper, ensuring quality control and maintaining standards \cite{kelly2014peerreview}. 
For example, code review helps programmers learn new implementation techniques while ensuring code quality for product deployment \cite{huang2016conflictpeerreview_CSCW16}. 
Similarly, peer review in research communities enables researchers to enhance science communication and promote the overall quality and trust within the research communities \cite{kelly2014peerreview}.

However, the effectiveness of peer feedback is often compromised by variability in the time and effort invested by the reviewers, especially since most of the peer review was done voluntarily.
Additionally, the tone of the feedback, which can occasionally be excessively negative, may impact the emotional state and task performance of the authors \cite{wu2018softenpain_CSCW18}. 
In response to these challenges, several strategies have been developed to improve the process of delivering and receiving peer feedback for both reviewers and authors.

To support reviewers, \citeauthor{cambre2018juxtapeer_CHI18} demonstrated that employing contrasting cases, which involve reviewers comparing two different submissions from different authors, improved the quality of peer feedback and engaged reviewers deeply \cite{cambre2018juxtapeer_CHI18}. 
Researchers also found that equipping reviewers with tools to quickly find examples to complement their feedback improved the quality of their critiques \cite{kang2018paragon}.
Additionally, it has been shown that rubrics \cite{yuan2016almostExpert} or structured scaffolding designed for novice reviewers facilitates the delivery of high-quality feedback during the peer review process \rhlp{\mbox{\cite{luther2015structuring_CSCW15, sun2024metawriter_CSCW24, greenberg2015critiki, hicks2016framing}}}. 
Moreover, encouraging reviewers to re-review their feedback based on the specific effective linguistic style, such as increasing specificity, before sending feedback \cite{krause2017critique} or to employ positive language, such as ``good job'' \cite{Nguyen2017fruitfulfeedback}, has proven beneficial in improving the perceived helpfulness and emotional responses of authors and their reception to critiques.

On the other hand, to support authors, it has been found that encouraging and scaffolding authors to reflect on their own creative work and prioritize which comments should be addressed first during the revision process can mitigate the negative impact when receiving low-quality peer feedback \cite{yen2017listen_CC17, yen2020decipher}. 
Additionally, some researchers have suggested employing various coping strategies, such as affirming one's self-worth before receiving critiques, having expressive writing, or diverting one's attention away from critique before revision, is effective in alleviating the negative effects when receiving critical feedback \cite{wu2018softenpain_CSCW18}.

Although various strategies have been proposed to enhance the peer review process, they often require additional effort from both reviewers and authors. 
Given that the creation and refinement of review content are essential to the peer review process and critical to the advancement of scientific knowledge, many argue that this responsibility should not be delegated to technology such as AI. 
Therefore, in this paper, we focus on how AI can assist in improving the tone of reviews. 
Specifically, we explore the effects of adding a positively reframed summary generated by AI, using Large Language Models (LLMs), based on the original critical feedback provided by reviewers.
Notably, the feedback itself remained unchanged, presented in its original form, with only the addition of a positive reframed summary at the end of the feedback.
We then examined how this adjustment was perceived by the authors.

\subsection{Cognitive Reframing for Positive Thinking}
\label{sec:LLM_reframing}
Receiving critical feedback has been found to elicit negative emotions in the peer review process \cite{wu2021betterFeedback_CSCW21}. 
These negative emotions can cause individuals to focus on negative aspects and feel discouraged, potentially hindering the iterative improvement of their work \cite{taggart2017affectMatters_2017, wu2021betterFeedback_CSCW21}. 
It has been shown that providing praise in feedback is crucial for reviewers to reinforce positive behavior and enhance the self-esteem of authors \cite{cavanaugh2013performanceFeedback_2013}, and it is often used to soften criticisms, thereby improving the relationship between reviewers and receivers \cite{hyland2001sugaringthepill_2001}. 

Similar to the function of praise, we see cognitive reframing, also known as cognitive reappraisal, has the potential to soften criticisms.
Cognitive reframing has been proven effective in various contexts. 
It can help in coping with anxiety during cognitive therapy \cite{clark2013cognitiverestructuring_2013}, reducing emotional distress when receiving supportive communication \cite{batenburg2014experimental}, and offering reframed social support to people with negative thinking patterns \cite{smith2021crowd-poweredcognitivereappraisal_CSCW21}. 
Cognitive reframing involves identifying and changing one's perspective on situations and thoughts, with the goal of altering their emotional impact \cite{troy2010cognitivereappraisalability}. 
However, effectively employing cognitive reframing is cognitively demanding, especially when under stress \cite{raio2013cognitiveregulation_PNAS}, and it usually requires adequate training or professional assistance \cite{troy2010cognitivereappraisalability}.

Therefore, we see the opportunity of using large language models (LLMs) to reframe critical feedback positively to support feedback reception.
With the development of LLMs, researchers have begun to apply LLMs to support cognitive reframing. 
For instance, a recent study demonstrated the potential of LLMs to guide individuals in reevaluating their situations and finding alternative ways to manage upsetting circumstances \cite{zhan2024LLMreappraisal}. 
A case study also revealed that LLMs can effectively help individuals overcome negative thinking and reduce emotional intensity \cite{sharma2024LLMrestructuring_CHI24}.

However, the current research primarily emphasizes the potential of LLMs in assisting cognitive reframing to manage mental health issues, with limited exploration of its impact on receiving critiques in peer feedback. 
As previously mentioned, receiving critical feedback can potentially lead authors to experience negative emotions \cite{kim2019PosNegFeedbackEmotion}, low self-efficacy \cite{mercer2024NegativeFeedbackHarms}, and low autonomy \cite{mspb_feedback_autonomy_meaningfulness}.
Hence, this study seeks to broaden the scope of LLMs' reframing capabilities to the peer review process. 
Particularly, we aim to investigate how LLMs can positively reframe critiques to potentially influence the way feedback is received.

\subsection{The Practice of Showing an Overall Evaluation in Peer Review}
\label{sec:overall_recommendation}
Research communities often adopt the practice of giving an overall evaluation in peer reviews to support reviewers in streamlining decision-making with other reviewers and guide authors to the expected next steps.
\rhl{To reviewers, the quantified score or categorization of recommendations provides them a rubric to assess the work and also facilitates reviewers and editors in making a consensus.
To authors, the quantified relative performance could influence their task performance and motivation. 
For instance, previous studies have shown that individuals who received randomly assigned high performance as positive feedback demonstrated better task performance compared to those who were randomly evaluated as low performers \cite{zhou1998feedback}.
Receiving scores also influences task motivation because individuals gauge their relative performance via scores by comparing themselves with others \cite{butler1987task}.   
Though presenting overall evaluation is a common practice in peer review, we lack an understanding of how presenting this relative performance influences authors' perceptions and their revision outcomes, especially when combined with an AI-reframed positive summary.}
As far as we know, the rationale for keeping this practice of giving an overall evaluation is often grounded in established editorial policies and general best practices \cite{chi2024_reviewing_guide, acm_tochi_reviewer_guidelines, vrst2023_reviewing, agu_review_criteria} rather than empirical studies.
Hence, we aim to explore how \textit{the presence of AI-reframed positive summary} combined with \textit{different overall evaluations} affects authors' critique reception.

\subsection{Attributional Theory of Motivation and Authors' Motivation for Making Revisions}
\label{sec:attributional_theory_of_motivation}
According to the \textit{attributional theory of motivation} \cite{Weiner2001, graham2020AttributionalTheoryOfMotivation}, the way authors attribute the cause of the feedback can also affect their attitude and subsequent revision behavior. 
The attributional theory of motivation \cite{Weiner2001}, first proposed by Bernard Weiner in the context of educational psychology, explains that people's motivation for a given task can be influenced by how they interpret the cause of success or failure. 
This includes their own responsibility (\textit{intra}personal attribution) or others' responsibility (\textit{inter}personal attribution). 
These two forms of attributions are not mutually exclusive, and the boundary between them can be unclear. 
However, this differentiation still enables researchers to better understand people's motivations. 
When individuals primarily adopt intrapersonal attribution, they tend to see the cause of their successes or failures as their own responsibility. 
This can lead to self-improvement or, if negative feedback is constantly received, to low self-efficacy \cite{Weiner2001}. 
Conversely, individuals who mainly adopt interpersonal attribution tend to attribute their learning outcomes to external factors, such as reviewers, feedback providers, instructors, teachers, or institutions. 
This can result in efforts to change external circumstances, such as the community, but may also lead to complaints about perceived unfairness within the community or society \cite{Weiner2001}.

Based on the intrapersonal and interpersonal factors in the attributional theory of motivation, it is expected that when authors receive critical feedback, they may work on improving themselves to enhance self-efficacy, or they may end up believing they are unable to succeed if they adopt \text{intra}personal attribution. 
Alternatively, authors may attempt to change external factors, such as the practices of their research community, or resort to complaining and possibly leaving the community if they mainly adopt \textit{inter}personal attribution.

Knowing how intrapersonal and interpersonal factors influence authors' motivation to make revisions enables researchers and designers to better provide support for peer feedback. 
Therefore, we also explored how authors describe their motivation to make revisions based on the critical feedback they received.

\section{Research Questions and Hypotheses}
\rhl{
In summary, the research questions were developed based on the \textit{attributional theory of motivation} by considering the effect of the \textit{presence of AI-reframed positive summary} and \textit{overall evaluation} on \textit{intrapersonal} (RQ1),  \textit{interpersonal} (RQ2) factors, and revision outcomes (RQ3).
We also examined potential motivational factors for authors to make revisions (RQ4).} 
\rhlp{Because there was not enough past work to inform hypotheses for the factor of \textit{overall evaluations}, we explored the impact of \textit{overall evaluations} on critique acceptance more broadly using only research questions.
Thus, we formed hypotheses for the factor of "presence of AI-reframed positive summary" but not for "overall evaluation".
}
Together, we asked the following four research questions:


\begin{itemize}
\item[\textbf{RQ1:}] \textbf{How does the combination of the presence of AI-reframed positive summary (present/absent) and different overall evaluations (high/low) affect the authors' intrapersonal perception when receiving critical feedback? }
\end{itemize}

\rhl{Under RQ1, we formed three hypothesis for the factor of \textit{presence of AI-reframed positive summary} based on Section \ref{sec:tech_supported_peer_review} and Section \ref{sec:LLM_reframing}.}


\begin{itemize}
    \item [] 
    \begin{itemize}
    \item[H1a:] Receiving an AI-reframed positive summary leads to more \textit{positive emotion} compared with not receiving an AI-reframed positive summary.
    \item[H1b:] Receiving an AI-reframed positive summary makes people have less decrement in their \textit{self-efficacy in writing} compared with not receiving an AI-reframed positive summary.
    \item[H1c:] People experience higher \textit{perceived autonomy and competence} when the critique comes with an AI-reframed positive summary compared with not receiving an AI-reframed positive summary.
    \end{itemize}
\end{itemize}

\begin{itemize}
\item[\textbf{RQ2:}] \textbf{How does the combination of the presence of AI-reframed positive summary (present/absent) and different overall evaluations (high/low) affect the authors’ interpersonal perception when receiving critical feedback?}
\end{itemize}

\rhl{Under RQ2, we formed two hypothesis based on Section \ref{sec:tech_supported_peer_review}.
We formed the following hypothesis for the factor of \textit{presence of AI-reframed positive summary} based on Section \ref{sec:tech_supported_peer_review} and Section \ref{sec:LLM_reframing}.}

\begin{itemize}
    \item [] 
    \begin{itemize}
    \item[H2a:] Receiving an AI-reframed positive summary increases the perceived fairness and usefulness of the critique.
    \item[H2b:] Receiving an AI-reframed positive summary makes people perceive the reviewer was fairer and more knowledgeable compared with not receiving it.
    \end{itemize}
\end{itemize}

\begin{itemize}
\item[\textbf{RQ3:}] \textbf{How does the combination of the presence of AI-reframed positive summary (present/absent) and different overall evaluations (high/low) influence the authors' revision outcomes when receiving critical feedback?}

\item[\textbf{RQ4:}] \textbf{What factors influence the authors' motivation to revise their writing when receiving critical feedback?}
\end{itemize}

\section{Method}
\subsection{Experiment Design}
To investigate the influence of showing \textit{AI-reframed positive summary} and \textit{overall evaluation} of the writing on authors' feedback reception, motivation, and revision outcome, we conducted a two-stage online controlled experiment, in which we asked participants to complete a writing task in stage 1 and then revise their writing based on the feedback they received in stage 2. 
In the writing task at stage 1, we instructed participants to write a 250-word cover letter for a job in social media marketing. 
We asked them to introduce themselves and describe how their characteristics, skills, and past experience make them suitable for this position. 
Then, we invited the same group of participants to join the study again three days later after completing their initial writing. 
We instructed participants to revise their initial writing based on the feedback we gave them at stage 2. 
We chose this writing task because we had to find a writing topic that was general enough for participants to complete in a relatively short period of time.
Also, the task should have no right or wrong answers that allow us to give participants a randomized overall evaluation. 

We manipulated and compared two types of feedback presentation (with and without AI-reframed positive summary) and two levels of overall evaluation (high-scored and low-scored overall ratings). This two-by-two design led to four conditions, including \textit{NoPosFramed-HighScored, PosFramed-HighScored, NoPosFramed-LowScored,} and \textit{PosFramed-LowScored} conditions. We randomly assigned participants to experience one of four conditions (between-subject design). 

\begin{table}[t]
\caption{Example of critical feedback and AI-reframed positive summary a participant received. \rhl{Participants saw the AI-reframed summary being appended to, not in place of, the critical feedback in the study.}}
\label{tab:example_feedback}
\resizebox{\columnwidth}{!}{%
\centering
\begin{tabular}{|>{\hspace{0pt}}m{0.669\linewidth}|>{\hspace{0pt}}m{0.273\linewidth}|} 
\hline
\textbf{Critical feedback} & \textbf{AI-reframed positive summary} \\ 
\hline
    1. Writing Style: The cover letter is excessively informal and lacks professional finesse.\par{}Starting with "To whom it may concern" is outdated, and the letter meanders without getting to the point. A sharp, engaging introduction tailored to the job and company is missing. \par\null\par{}2. Professional Standing: Mentioning unrelated degrees (Marine Biology and Neurobiology) without effectively tying these to the required social media content creation skills shows a fundamental misunderstanding of the job's nature. The attempt to link these degrees with organizational skills is weak and unpersuasive.\par\null\par{}3. Work Experience: Your experience in sales and as an environmental educator, while interesting, is only tangentially relevant. The crucial omission is your lack of direct experience in social media content creation or using analytics tools professionally. Asserting that you believe you can easily learn does not substitute for proven experience.\par\null\par{}4. Key Accomplishments and Relevance: There's no indication of proficiency in graphic design tools, strong visual aesthetics, or specific accomplishments in content creation that would be directly relevant to this role. Simply stating personal use of social media and a vague familiarity with technology trends falls significantly short of demonstrating the expertise or passion for social media content creation this position demands. & Dear Applicant,\par{}Your cover letter is a promising foundation for showcasing your talents. \par{}A more dynamic opening could better reflect your creativity and personalizing it to the hiring manager shows you're truly invested.\par{}Highlighting specific achievements and providing concrete examples of your success will not only substantiate your claims but also bring your expertise to life. \par{}Don't forget to detail your experience with social media platforms, emphasizing how your skills meet the job's visual and design demands. \par{}This feedback is a splendid opportunity to refine your application into one that not only meets but exceeds expectations. \par\null\par{}Best regards, \\
\hline
\end{tabular}
}
\end{table}

In \textit{NoPosFramed-HighScored} condition, we provided critical feedback based on participants' writing at stage 1. 
To maintain the consistency of the feedback we gave to participants in terms of the aspects of criticism, we used the GPT4 model\footnote{We used the gpt-4-0125-preview model for generating critical feedback.} 
to generate critical feedback, specifically focusing on the authors' \textit{writing style, professional standing, work experience, and key accomplishments and their relevance} in applying for a specific job position in the writing task. 
As the response of the GPT models was designed to be expressed as positively as possible \cite{openai_gpt4_system_card}, we explicitly mentioned \textit{being a strict reviewer} and \textit{using a harsh tone} in the prompt for generating critical feedback. 
The detailed prompt can be referred to from the supplementary document.
Please note that we did not inform participants about the source of their feedback. 
However, we asked them to identify who they thought might have provided the feedback, including ChatGPT, another crowd worker, a social media expert, or a member of our research team.
26\% of participants (n=26) reported that they believed ChatGPT was the author of the feedback.
In addition to providing customized feedback to participants, we also showed them an overall evaluation prior to the feedback, saying, ``The score of your cover letter is 4 out of 5 among all candidates for this position.'' The purpose was to make people perceive that they received a high evaluation of their writing.   

In \textit{PosFramed-HighScored} condition, participants saw a positively reframed summary at the end of the critical feedback they received. 
We also used the same GPT4 model to generate a positively reframed summary based on the customized feedback we prepared for each participant. 
We followed the practice of encouraging cognitive reframing in cognitive therapy \cite{psychiatric_journal_article} to design the prompt by making the GPT4 model interpret the reviewers' intention by thinking about a benefit or upside to a negative situation using a positive tone. 
Also, the reframed content can involve identifying a lesson to be learned from a difficult situation. 
The example of critical feedback and AI-reframed positive summary is in \autoref{tab:example_feedback}.
The purpose of showing a positively reframed summary was to encourage authors to interpret the critical feedback in a positive and constructive way regardless of the content and tone of the feedback they received. 
In addition to providing customized feedback with a positively reframed summary to participants, we also showed them the same overall evaluation as we did in \textit{NoPosFramed-HighScored} condition. 
The detailed prompt we used for generating the positively reframed summary can be referred to from the supplementary document. 

In \textit{NoPosFramed-LowScored} condition and \textit{PosFramed-LowScored} condition, we followed the same approach to generate critical feedback and positively reframed summary for every participant, depending on their condition. The only difference was the overall evaluation we showed participants prior to the feedback, which said, ``The score of your cover letter is 2 out of 5 among all candidates for this position.'' The purpose was to make people perceive that they received a low evaluation of their writing.


\subsection{Participants}
We recruited 137 participants (70 females, 63 males, 4 unspecified) with a mean age of 29.75 years old (\textit{SD} = 8.81) from Prolific\footnote{We recruited participants from https:\//www.prolific.com/, an online participant recruiting and data collection platform}. 
All participants were paid \$10.5 GBP for participating in the approximately 65-minute study. 
The payment was determined based on Prolific's standard. 
We recruited participants by telling them this study required them to join two times by writing a cover letter for applying for a pseudo-job position (stage 1) and revising their writing a few days later (stage 2). 
The ethical review board of the authors’ institute approved the study.

\begin{figure*}[t]
  \centering
  \includegraphics[width=0.8\linewidth]{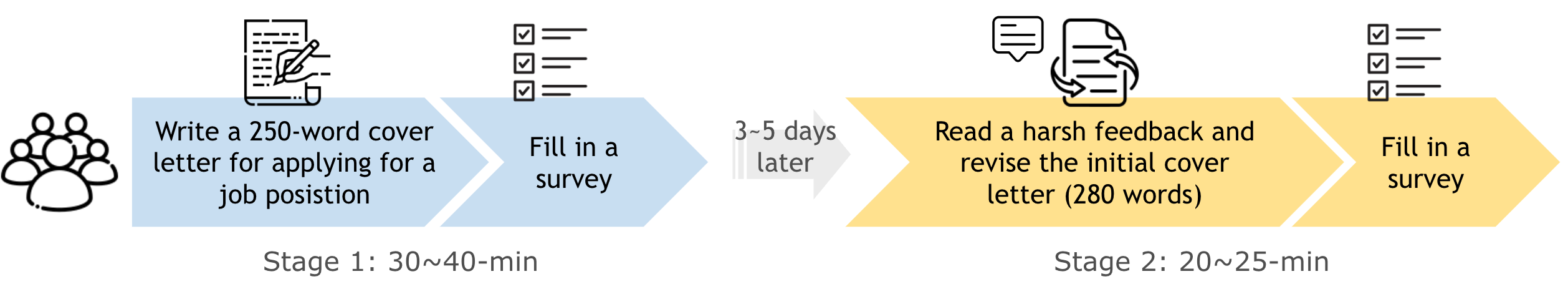}
  \caption{Procedure of the controlled online experiment.}
  \Description{temp}
  \label{fig:procedure}
\end{figure*}

\begin{table}[t]
\caption{Experiment design and sample size for each condition.}
\label{tab:conditions}
\resizebox{\linewidth}{!}{%
\begin{tabular}{|l|l|l|}
\hline
& {\color[HTML]{1C4587} \begin{tabular}[c]{@{}l@{}}critical feedback only \\ (baseline, 150$\sim$200 words)\end{tabular}} & {\color[HTML]{1C4587} \begin{tabular}[c]{@{}l@{}}critical feedback (150$\sim$200 words) + \\ AI-reframed positive summary (50 words)\end{tabular}} \\ \hline
{\color[HTML]{986536} See high overall evaluation} & {\color[HTML]{595959} \begin{tabular}[c]{@{}l@{}}Score: 4/5\\ (N = 31)\end{tabular}}                                   & {\color[HTML]{595959} \begin{tabular}[c]{@{}l@{}}Score: 4/5 + AI-reframed \\ positive summary\\ (N = 37)\end{tabular}}                            \\ \hline
{\color[HTML]{986536} See low overall evaluation}  & {\color[HTML]{595959} \begin{tabular}[c]{@{}l@{}}Score: 2/5 \\ (N = 32)\end{tabular}}                                  & {\color[HTML]{595959} \begin{tabular}[c]{@{}l@{}}Score: 2/5 + AI-reframed \\ positive summary summary\\ (N = 37)\end{tabular}}                    \\ \hline
\end{tabular}
}
\end{table}

\subsection{Procedure}
The study was carried out on a website built with Google Cloud Firestore. 
In stage 1, participants joined the study directly via Prolific and accessed the Google form to read the pseudo-job description, which was a social media content creator for a tech company (Please refer to the details of the task instructions from the supplementary document). 
We instructed participants to fill out the Google form with their demographic information and basic information related to the job description, including their work experience and skills. The purpose was to engage participants by simulating the process of filling in an application form in an actual job application process. Next, we redirected participants to the website, where they were instructed to write a 250-word cover letter based on their work experience and skills, which they had just filled out on the Google form. After finishing the writing task, participants ended stage 1 by completing a survey via Google form. This process took approximately 30 to 40 minutes, depending on the participants' writing speed. 

Approximately three to seven days later, we invited the same group of participants to join the study again. Participants would see the feedback with an overall evaluation depending on the condition to which they were randomly assigned. Afterward, they were presented with their initial writing and then instructed to revise it based on the feedback they received. In this stage, we asked participants to revise their cover letters to at least 280 words (from 250). This word limit was intended to encourage participants to make some revisions at least. Once participants finished revising their cover letters, they filled in another Google form to end the study. \rhl{At the end of the survey, we debriefed participants that the critical feedback and reframed summary were both generated by AI.} This process took approximately 20 to 25 minutes, depending on the participants' revising speed (\autoref{fig:procedure}). 

For context, we did not forbid participants from using generative AI in the study to replicate real-life scenarios where people might use assistive technology for writing tasks. 
However, we disabled the copy-and-paste function in our web app to prevent direct copy-pasting. 
In the end, 29.2\% of the participants (n = 40) reported using generative AI during the writing process.

\subsection{Measurements}

\begin{table*}[h!]
\caption{\rhl{Rubric for evaluating revision quality of the cover letter. Two independent raters, who were unaware of the research purpose, evaluated all 137 revisions based on this rubric. The maximum score for each revision is 20, while the minimum score is 4.}}
\label{tab:rubric}
\resizebox{0.8\textwidth}{!}{%
\begin{tabular}{|l|l|l|l|}
\hline
 &
  \textbf{5-Effective} &
  \textbf{3-Average} &
  \textbf{1-Need improvement} \\ \hline
\textbf{Format: grammar} &
  \begin{tabular}[c]{@{}l@{}}No spelling, punctuation or \\ grammar errors\end{tabular} &
  \begin{tabular}[c]{@{}l@{}}Some spelling or grammatical \\ errors found\end{tabular} &
  \begin{tabular}[c]{@{}l@{}}Many errors that take \\ focus away from content\end{tabular} \\ \hline
\textbf{Format: basic information} &
  \begin{tabular}[c]{@{}l@{}}Covers basic information but \\ offers an engaging and gripping \\ way into the body of the letter \\ and clearly connects the person \\ to the position\end{tabular} &
  \begin{tabular}[c]{@{}l@{}}Covers basic information but \\ only a lackluster way of getting \\ into the core content of the letter\end{tabular} &
  \begin{tabular}[c]{@{}l@{}}May or may not cover basic \\ information and only a tenuous \\ or weak way into the body of \\ the letter and establishes no link \\ between the person and the position\end{tabular} \\ \hline
\textbf{Content: skills} &
  \begin{tabular}[c]{@{}l@{}}Emphasizes skills or abilities the \\ person has that relate to the job \\ for which they are applying. \\ Mentions work, volunteer, and \\ education experiences.\end{tabular} &
  \begin{tabular}[c]{@{}l@{}}Limited information on skills or \\ abilities the person has that relate \\ to the job they are applying for. \\ Some unsupported assertions but \\ also some good examples of the \\ connections between the person \\ and the position\end{tabular} &
  \begin{tabular}[c]{@{}l@{}}List of Skills (i.e.- Communication, \\ Flexibility, and Teamwork) with no \\ evidence of work, volunteer, or \\ education experience\end{tabular} \\ \hline
\textbf{Content: qualifications} &
  \begin{tabular}[c]{@{}l@{}}Identifies one or two of the person's \\ strongest qualifications and clearly \\ relates how these apply to the job \\ at hand.\end{tabular} &
  \begin{tabular}[c]{@{}l@{}}Identifies one of the person's \\ qualifications, but it is not \\ related to the position at hand.\end{tabular} &
  \begin{tabular}[c]{@{}l@{}}Does not discuss any relevant qualifications. \\ Have not related the person's skills to the \\ position applied for.\end{tabular} \\ \hline
\end{tabular}%
}
\end{table*}

\subsubsection{Perceived Emotional Valence of the Feedback (H1a)}
To examine how participants feel after receiving critical feedback, we asked them to indicate their feelings based on the following question adapted from \rhl{\mbox{\citeauthor{rasmussen2009emotional}}} \cite{rasmussen2009emotional}: ``The feedback I received, as I recall are: 1-extremely negative, and 7-extremely positive''

\subsubsection{Change of Self-Efficacy in Writing (H1b)}
We also evaluated participants' self-efficacy before and after receiving critical feedback with a 7-point Likert scale, ranging from 1 (strongly disagree) to 7 (strongly agree). 
The nine items for self-efficacy were adapted from \rhl{\mbox{\citeauthor{mitchell2017exploring}}} \cite{mitchell2017exploring}.
\rhl{The original scale was developed for evaluating people's self-efficacy for academic writing. We changed a few keywords and replaced them with writing a cover letter and self-introduction for applying for a job to fit the study context. Example items include} \textit{\rhl{``I feel I have the skills to write a cover letter (self-introduction) for this job application''}}, \textit{\rhl{``Writing a cover letter (self-introduction) for this job application comes easily to me.''}}
The Cronbach’s Alpha for the survey was $\alpha$ = .918. 
We calculated the average score from nine items from stage 1 and stage 2. 
The difference between the two scores from each stage was calculated to represent participants' change in self-efficacy.

\subsubsection{Perceived Autonomy and Competence (H1c)}
\rhl{Regarding how feedback influenced perceived autonomy and competence, we adapted the questions from \mbox{\citeauthor{sheldon2001satisfying}} \cite{sheldon2001satisfying}. 
Participants answered the following three items for autonomy with a 7-point Likert scale, ranging from 1 (not at all) to 7 (very much):} \textit{When reading the feedback and revising the text, I felt...that my choices were based on my true interests and values; free to do things my own way; that my choices expressed my ``true self'';}, \rhl{and another three items for competence: }\textit{When reading the feedback and revising the text, I felt...that I was successfully completing difficult tasks and projects; that I was taking on and mastering hard challenges; very capable in what I did.} 
\rhl{The Cronbach’s Alpha for the three autonomy items was $\alpha$ = .756, and for competence, it was $\alpha$ = .804.
We averaged the scores from three items for each construct, representing participants' perceived autonomy and competence, respectively.}

\subsubsection{Perception of the Peer Feedback (H2a)}
To examine how much participants accepted the feedback they received, we adapted the survey from \rhl{\mbox{\citeauthor{Nguyen2017fruitfulfeedback}}} \cite{Nguyen2017fruitfulfeedback} and asked them to rate the perceived usefulness and fairness of the feedback. 
The questions related to perceived usefulness include, \textit{The feedback I received influenced my revision of the work; I applied the feedback I received when I edited my work.} Whereas the questions related to perceived fairness include, \textit{The feedback I received provides helpful suggestions for improving my work; The feedback I received is constructive; The feedback I received is fair; The feedback I received is relevant to my work.} 
\rhl{The Cronbach’s Alpha for the three items in \textit{perceived usefulness} was $\alpha$ = .827, while the Cronbach’s Alpha for the three items in \textit{fairness of the feedback} was $\alpha$ = .911.}
Participants answered all survey items with a 7-point Likert scale, ranging from 1 (strongly disagree) to 7 (strongly agree). 
We averaged the scores from all items from the sub-scale and used each of them to represent the \textit{perceived usefulness} and \textit{fairness of the feedback}, respectively.

\subsubsection{Perception of the reviewer (H2b)}
We also adapted the survey from \rhl{\mbox{\citeauthor{Nguyen2017fruitfulfeedback}}} 
\cite{Nguyen2017fruitfulfeedback} to assess how participants perceived the reviewers, including their fairness and expertise. 
The questions regarding the fairness of reviewers were: \textit{The person who gave me feedback was considerate; The person who gave me feedback was sincere; The person who gave me feedback was polite; The person who gave me feedback was disrespectful \footnote{This is a reversed item}.} 
The questions regarding the expertise of reviewers were: \textit{The person who gave me feedback was qualified; The person who gave me feedback was an expert in the task; The person who gave me feedback was knowledgeable.} 
\rhl{The Cronbach’s Alpha for the three items in} \textit{\rhl{perceived usefulness of the reviewer}}\rhl{ was $\alpha$ = .87, while the Cronbach’s Alpha for the three items in }\textit{\rhl{expertise of the reviewers}}\rhl{ was $\alpha$ = .936.}
These items were also measured with a 7-point Likert scale, where one indicated strongly disagree, and seven indicated strongly agree. 
We averaged the scores from all items from the sub-scale and used each of them to represent the \textit{perceived fairness} and \textit{expertise of the reviewers}, respectively.

\subsubsection{Revision Outcome (RQ3)}
\label{sec:revision_outcome}
In addition to their perception, we also measured their actual revision outcomes, including quantity and quality of revision. 

Regarding the quantity of revision, we compared the document similarity between participants' initial writing and their final revision using cosine similarity, which is one of the common approaches for calculating semantic similarity between two documents \cite{datta2017identifying, rho2018fostering, haring2018addressed}.
\rhl{We first converted text into numerical vectors using TF-IDF (Term Frequency-Inverse Document Frequency) and then calculated the cosine similarity between the two vectors. We used TF-IDF because this captured word-level information, and the weights assigned to words are relatively easy to interpret \footnote{\rhl{We also tried BERT embeddings, which captured rich semantic relationships between words and contexts, with cosine similarity and got a similar statistical result. The conclusion of the revision quantity remained the same. We put the result in the supplementary file.}}.}
The cosine value ranges from 0 to 1.
If the angle between two vector representations is small (close to 0 degrees), the cosine value is close to 1, meaning the two documents are very similar. 
If the angle is large (close to 90 degrees), the cosine value is close to 0, meaning the documents are not similar.
In other words, a high cosine value indicates participants made few semantic revisions. 

\rhl{Regarding the quality of revision, we developed a rubric by synthesizing the rubrics we found from the career center that taught students how to write a cover letter \cite{wcu_cover_letter_rubric, humboldt_cover_letter_rubric, indstate_cover_letter_rubric}.
The rubric consisted of four main aspects: grammar, basic information about the applicant, skills mentioned by the applicant, and qualifications (\autoref{tab:rubric}). A 5-point Likert scale was used to score each item, ranging from ``needs improvement (1)'' to ``effective (5).'' This allowed for a maximum score of 20 for each revision. 
We then recruited two external raters who were unaware of the research purpose to evaluate all 137 revisions based on the rubric.
We paid each rater 78.15 USD (this is slightly above the hourly wage of the country where the study was conducted) for approximately five to six hours of work to assess all revisions.
The averaged scores from two raters were analyzed as an index for quality of revision.
}


In addition to the behavioral index, we also collected participants' perceptions about their revision quality.
We asked participants the following three questions with a 7-point Likert scale, ranging from 1 (low) to 7 (high): 
\textit{How much effort did you invest in the writing task?
How would you rate the quality of the writing task?
How confident are you that the writing task fully satisfied the goals?}
Three items were adapted from \cite{yen2017listen_CC17}, and the Cronbach’s Alpha for the three items was $\alpha$ = .858.
The average score served as an index for participants' subjective assessment of their revision quality.

\begin{figure*}[h!]
  \centering
  \includegraphics[width=0.7\linewidth]{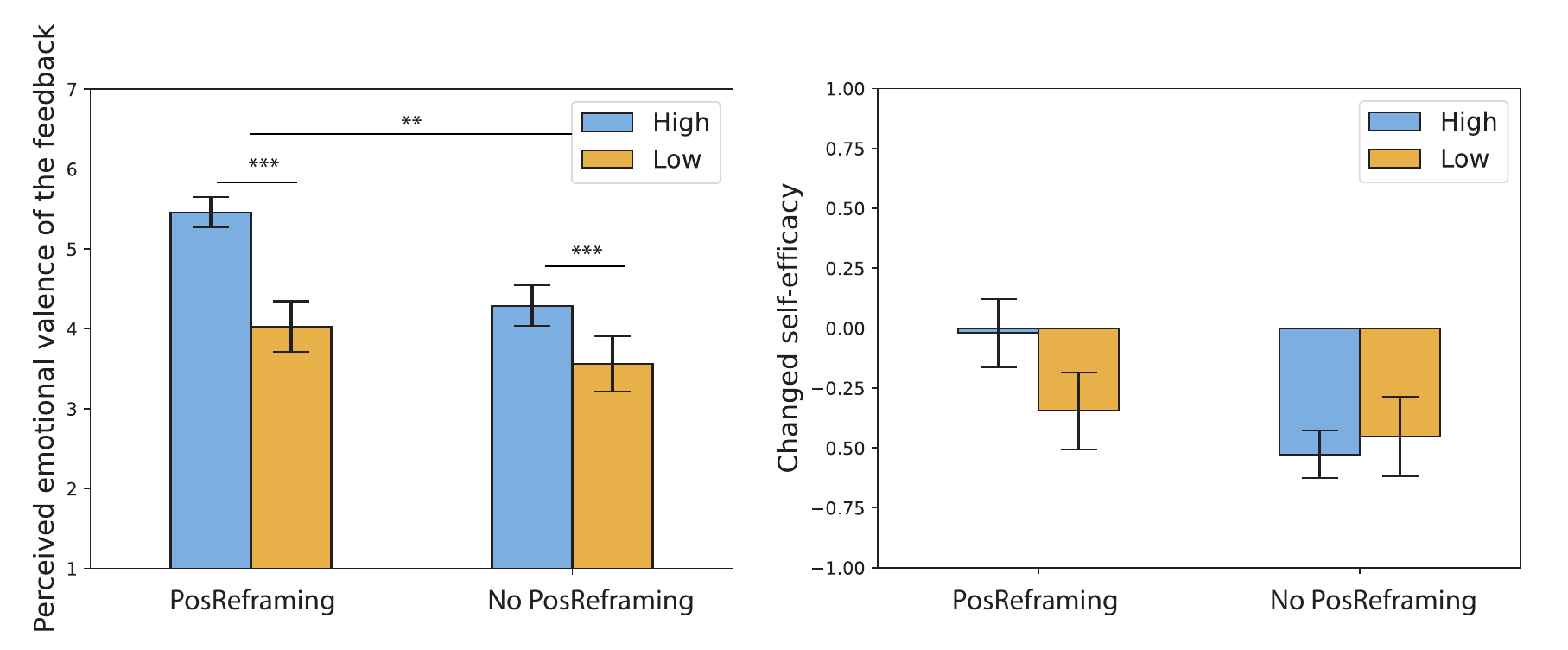}
  \caption{\rhl{Mean ($\pm$ standard error of the mean) of the perceived emotional valence of the feedback (Left; H1a) and the change in self-efficacy between initial writing and revision (Right; H1b). The X-axis shows whether participants received \textit{AI-reframed positive summary}. The Y-axis of the left figure shows the perceived emotional valence. A high score indicates participants perceived positive emotion. The Y-axis of the right figure shows the change in self-efficacy. A score closer to zero indicates a smaller decrease in self-efficacy.} (** indicates $p<.01$, *** indicates $p<.001$)}
  \Description{temp}
  \label{fig:emotion and self-efficacy}
\end{figure*}

\subsubsection{Potential Factors that Influence One's Motivation to Make Revision (RQ4)}
To investigate what factors determine participants' decision to make revisions, we asked participants to elaborate on their reasons for revising or not revising their cover letters in the open-ended responses of the survey.
\rhl{The data from participants' open-ended comments enabled us to identify participants' ``claimed motivation''.}
We followed the thematic analysis method~\cite{braun_2006_thematicanalysis} to analyze the open-ended responses. One of the authors open-coded all relevant concepts, assigned labels that featured the concepts, and grouped labels into different themes. 
Next, all the authors repeatedly discussed the quotes within each theme and the themes themselves. 
Finally, the developed themes were compared and adapted among all participants until they covered the data exhaustively.

\section{Results}
In the following analysis, we first tested the homogeneity of variances across groups using Levene's test for each of our dependent variables. 
We also tested the normality of the distribution of each dependent variable using the Shapiro-Wilk test. 
If both assumptions were met (i.e., the data showed homogeneity of variances according to Levene's test and normality according to the Shapiro-Wilk test), we proceeded with the standard two-way ANOVA with two independent variables, \textit{with/without AI-reframed summary} and \textit{high/low overall evaluation}, to analyze the data.
If one or both assumptions were violated (i.e., if Levene's test indicated unequal variances or the Shapiro-Wilk test suggested non-normality), we applied the Aligned Rank Transform (ART) \cite{wobbrock2011aligned} to the dependent variables first. 
After applying ART, we then conducted the two-way ANOVA on the transformed data.

\subsection{Effect on Intrapersonal Perception (RQ1)}
\label{sec:RQ1}
To investigate the effect of \textit{AI-reframed positive summary} and \textit{overall feedback evaluation} on receivers' perception of themselves (RQ1), we compared their emotion, self-efficacy for writing, and perceived autonomy and competence across four conditions.

\subsubsection{Positively Reframed Summary and High Scored Feedback Induced Positive Emotion (H1a)}
\label{sec:H1a}
The result of a two-way ANOVA using aligned ranked transformation (ART) revealed that there was no significant interaction effect of \textit{AI-reframed positive summary} and \textit{overall feedback evaluation} on participants' emotional valence ($F[1, 133]=0.97$, $n.s.$).
However, we found a significant main effect for both \textit{AI-reframed positive summary} and \textit{overall feedback evaluation} on participants' emotional valence.
Regardless of receiving a high or low overall evaluation, participants had significantly more positive emotions when receiving AI-reframed positive summary than the original critical feedback (\rhl{\textbf{H1a was supported},} $F[1, 133]=7.71$, $p<.01$, $\eta^2_p=.054$; PosFramed: $M=4.74$, $SD=1.55$; NoPosFramed: $M=3.92$, $SD=1.89$). 
Regardless of receiving AI-reframed positive summary or not, participants had significantly more positive emotions when receiving a high overall evaluation than a low overall evaluation of the feedback ($F[1, 133]=14.95$, $p<.001$, $\eta^2_p=.10$; HighScored: $M=4.93$, $SD=1.58$; LowScored: $M=3.81$, $SD=1.76$). 
(\autoref{fig:emotion and self-efficacy}, left)

Receiving \textit{AI-reframed positive summary} and \textit{a high overall evaluation} of the feedback led people to experience a more positive emotion than without receiving \textit{AI-reframed positive summary} and \textit{a low overall evaluation} of the feedback. 


\subsubsection{Reduced Self-Efficacy in Writing After Receiving Critical Feedback (H1b)}
The result of a two-way ANOVA with ART revealed that there was no significant interaction effect of \textit{AI-reframed positive summary} and \textit{overall feedback evaluation} on participants' change of self-efficacy in writing ($F[1, 133]=2.90$, $n.s.$).
We also did not observe a significant main effect of either \textit{AI-reframed positive summary} (\rhl{\textbf{H1b was not supported}}) or \textit{overall feedback evaluation} on participants' change of self-efficacy in writing, though we observed the trend that all participants reduced self-efficacy in writing after receiving critiques. 
(\autoref{fig:emotion and self-efficacy}, right)

Regardless of the presence of \textit{AI-reframed positive summary} and the score of \textit{overall feedback evaluation}, participants experienced a decrease in self-efficacy after receiving critical feedback.

\subsubsection{Positively Reframed Summary Induced High Autonomy and Competence (H1c)}
\begin{figure*}[h!]
  \centering
  \includegraphics[width=0.7\linewidth]{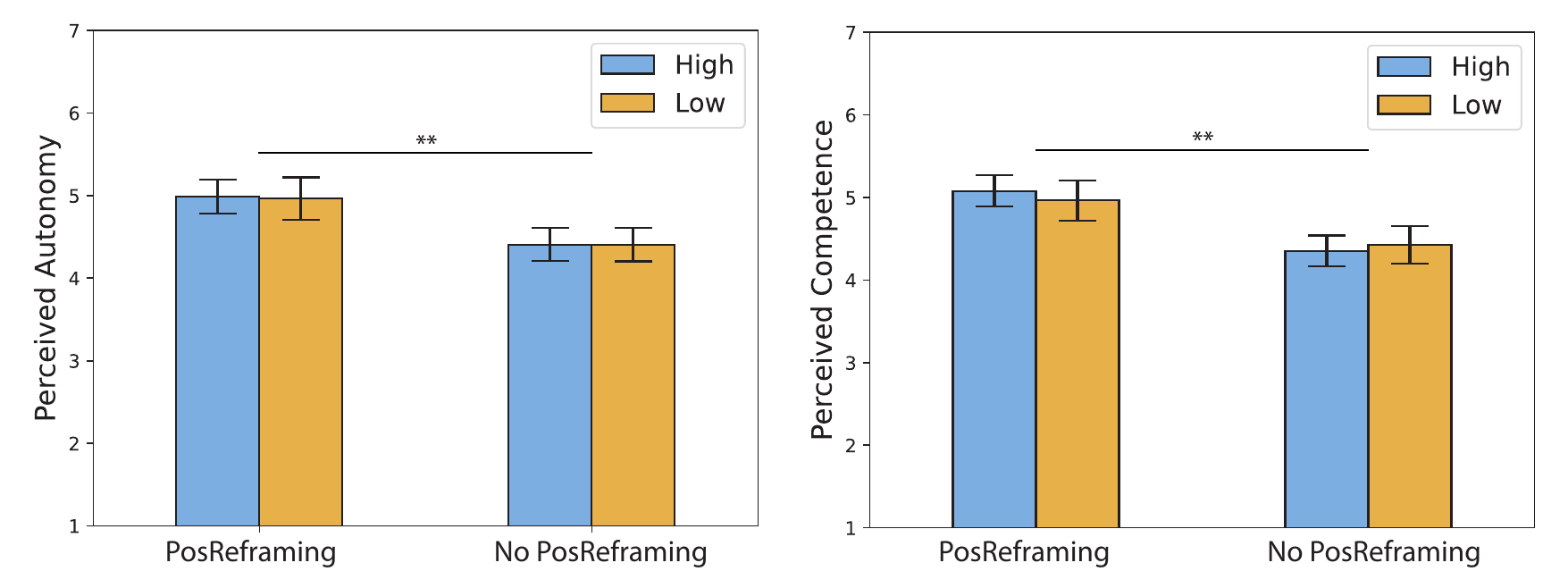}
  \caption{\rhl{Two bar charts with a mean ($\pm$ standard error of the mean) of participants' perceived autonomy (left) and competence (right) in making revisions.} 
  The X-axis shows the factor of the presence of \textit{AI-reframed positive summary}. The Y-axis shows participants' perceived autonomy and competence. A higher score indicates a higher perceived autonomy and competence (** indicates $p<.01$).}
  \Description{TEMP.}
  \label{fig:Autonomy_and_Competence}
\end{figure*}

The result of a two-way ANOVA revealed that there was no significant interaction effect of \textit{AI-reframed positive summary} and \textit{overall feedback evaluation} on \rhl{participants' perceived autonomy ($F[1, 133]=0.003$, $n.s.$) and competence ($F[1, 133]=0.21$, $n.s.$) in making revisions.}
However, there was a significant main effect for \textit{AI-reframed positive summary} on participants' perceived autonomy and competence.
Regardless of receiving a high or low overall evaluation, participants experienced higher autonomy and competence for revision when the critical feedback came with an AI-reframed positive summary, compared with the original critical feedback (\rhl{\textbf{H1c was supported}. Autonomy: $F[1, 133]=6.95$, $p<.01$, $\eta^2=.05$; PosFramed: $M=4.98$, $SD=1.22$; NoPosFramed: $M=4.41$, $SD=1.29$; Competence: $F[1, 133]=9.12$, $p<.01$, $\eta^2=.06$; PosFramed: $M=5.02$, $SD=1.13$; NoPosFramed: $M=4.39$, $SD=1.3$). 
(\autoref{fig:Autonomy_and_Competence})}

Receiving critical feedback came with an \textit{AI-reframed positive summary} significantly increased people’s perceived autonomy and competence in making revisions, regardless of the score of the overall evaluation. 

\subsection{Effect on Interpersonal Perception (RQ2)}
\label{sec:RQ2}
Next, we moved on to investigate the effect of \textit{AI-reframed positive summary} and \textit{overall feedback evaluation} on receivers' perception of the feedback they received (RQ2). 
We compared their perception of the feedback itself and the reviewers across four conditions.

\subsubsection{Positively Reframed Summary Increased Perceived Fairness of the Peer Feedback (H2a)}
\label{sec:H2a}
We asked participants to rate the usefulness and fairness of the feedback they received.
The result of a two-way ANOVA with ART revealed that there was no significant interaction effect of \textit{AI-reframed positive summary} and \textit{overall feedback evaluation} on neither the usefulness ($F[1, 133]=0.07$, $n.s.$) nor fairness ($F[1, 133]=0.005$, $n.s.$) of the feedback.
However, we found the main effect for \textit{AI-reframed positive summary} on participants' perceived fairness of the feedback.
Regardless of receiving a high or low overall evaluation, participants thought the critical feedback was fairer when it came with an AI-reframed positive summary, compared with the original critical feedback (\rhl{\textbf{H2a was partially supported}}. $F[1, 133]=8.62$, $p<.01$, $\eta^2_p=.06$; PosFramed: $M=6.17$, $SD=0.85$; NoPosFramed: $M=5.45$, $SD=1.47$).
The above effect was not found for participants' perceived usefulness of the feedback . 
(\autoref{fig:perception of feedback}A and B)

Receiving \textit{AI-reframed positive summary} increased the perceived fairness of the critical feedback, but did not impact the usefulness of the critical feedback. 

\subsubsection{Positively Reframed Summary Increased Perceived Fairness of the reviewers (H2b)}
\label{sec:H2b}
\begin{figure*}[h!]
  \centering
  \includegraphics[width=0.7\linewidth]{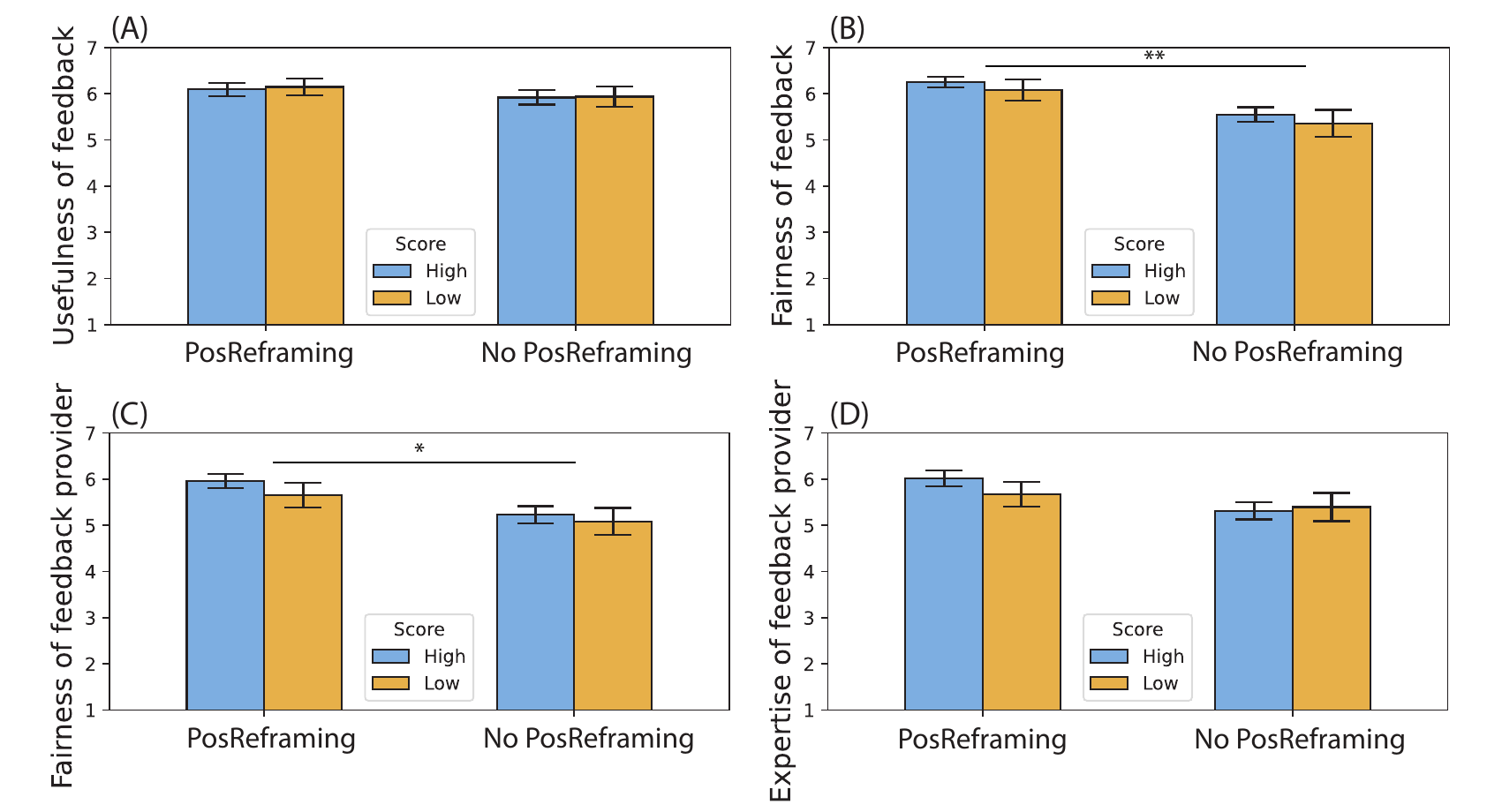}
  \caption{\rhl{Bar charts with a mean ($\pm$ standard error of the mean) of participants' perceived usefulness (left) and fairness (right) of the feedback (A, B), and fairness and expertise of the reviewer (C, D). The X-axis shows the factor of the presence of \textit{AI-reframed positive summary}. The Y-axis shows participants' perceived usefulness and fairness of the feedback. A higher score indicates higher perceived usefulness (A), fairness (B, C), and expertise (D) of the feedback and reviewers (* indicates $p<.05$, ** indicates $p<.01$).}}
  \Description{TEMP.}
  \label{fig:perception of feedback}
\end{figure*}

Next, we asked participants to rate the expertise and fairness of the reviewers.
The result of a two-way ANOVA with ART revealed that there was no significant interaction effect of \textit{AI-reframed positive summary} and \textit{overall feedback evaluation} on neither the expertise ($F[1, 133]=1.40$, $n.s.$) nor fairness ($F[1, 133]=0.20$, $n.s.$) of the reviewers.
However, we found the main effect for \textit{AI-reframed positive summary} on participants' perceived fairness of the reviewers.
Regardless of receiving a high or low overall evaluation, participants thought the reviewers were fairer when the critical feedback came with an AI-reframed positive summary, compared with the original critical feedback (\rhl{\textbf{H2b was partially supported}}. $F[1, 133]=4.962$, $p<.05$, $\eta^2_p=.04$; PosFramed: $M=5.8$, $SD=1.05$; NoPosFramed: $M=5.15$, $SD=1.56$).
A similar effect was not found for participants' perceived expertise with the reviewers. 
(\autoref{fig:perception of feedback} C and D)

Receiving \textit{AI-reframed positive summary} increased participants' perceived fairness of the reviewer, but did not impact the perceived expertise of the reviewer.

\subsection{Effect on Revision Outcome (RQ3)}
\label{sec:RQ3}
\subsubsection{Low Scored Feedback Increased Quantity of Revision}
We calculated the cosine similarity \rhl{using TF-IDF} between participants' initial writing and their final revision outcomes. 
This index indicates the quantity of participants' revisions.
The result of a two-way ANOVA with ART revealed that there was no significant interaction effect of \textit{AI-reframed positive summary} and \textit{overall feedback evaluation} on participants' quantity of revision ($F[1, 133]=0.002$, $n.s.$).
However, we found the main effect for \textit{receiving overall evaluation} on participants' quantity of revision.
Regardless of receiving an AI-reframed positive summary or not, participants made more revisions when the critical feedback was presented with a low-scored overall evaluation, compared with a high-scored overall evaluation ($F[1, 133]=5.36$, $p<.05$, $\eta^2_p=.04$; LowScored: $M=0.81$, $SD=0.13$; HighScored: $M=0.87$, $SD=0.11$). 
(\autoref{fig:similarity_and_perceived_quality}, Left)

\subsubsection{Quality of Revision}
\rhl{
The quality of the revision was the average score of two independent raters who evaluated the revisions based on a rubric we developed. 
The score for the quality of revision ranged from 4 to 20. 
The higher the score, the higher the quality of the revision.
The result of a two-way ANOVA revealed that there was no significant interaction effect of \textit{AI-reframed positive summary} and \textit{overall feedback evaluation} on participants' quality of revision ($F[1, 133]=0.043$, $n.s.$).
We also did not find any main effect of \textit{AI-reframed positive summary} ($F[1, 133]=0.12$, $n.s.$) and \textit{overall feedback evaluation} ($F[1, 133]=0.65$, $n.s.$) on participants' quality of revision.
(PosFramed-HighScored: $M=13.15$, $SD=3.15$; PosFramed-LowScored: $M=13.45$, $SD=2.54$; NoPosFramed-HighScored: $M=12.87$, $SD=2.73$; NoPosFramed-LowScored: $M=13.38$, $SD=3.16$)
}

\begin{figure*}[t]
  \centering
  \includegraphics[width=0.8\linewidth]{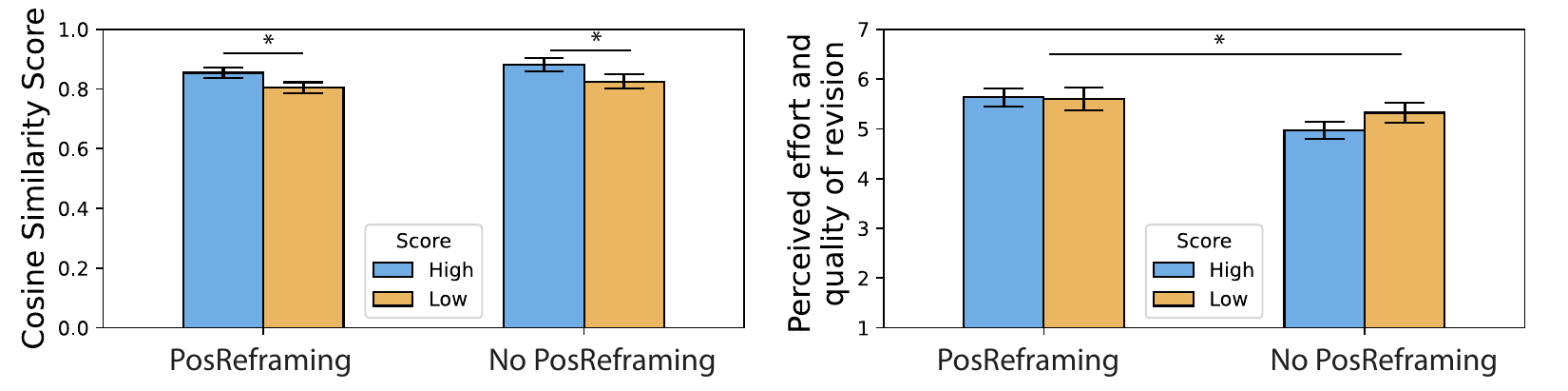}
  \caption{Two bar charts with a mean ($\pm$ standard error of the mean) of cosine similarity score for the revision outcome (left) and perceived effort and quality of revision (right). The X-axis shows the factor of the presence of \textit{AI-reframed positive summary}. The Y-axis for the left figure shows the cosine similarity score, ranging from zero to one. A score closer to one indicates a larger difference between the initial writing and final revisions. The Y-axis for the right figure shows participants' perceived effort and quality of revision. A higher score indicates increased effort and quality for the revision (* indicates $p<.05$).}
  \Description{TEMP.}
  \label{fig:similarity_and_perceived_quality}
\end{figure*}

However, we also evaluated participants' perception of their revision quality (Section \ref{sec:revision_outcome}) and found that \textit{AI-reframed positive summary} significantly influenced participants' perception of the effort and quality of their revisions.
The result of a two-way ANOVA with ART showed that regardless of receiving a high or low overall evaluation, participants perceived their effort and the revision quality was significantly higher when the critical feedback came with an AI-reframed positive summary, compared with the original critical feedback ($F[1, 133]=5.56$, $p<.05$, $\eta^2_p=.04$; PosFramed: $M=5.62$, $SD=1.08$; NoPosFramed: $M=5.15$, $SD=1.21$). 
(\autoref{fig:similarity_and_perceived_quality}, right)

Taken together, to answer RQ3, our result suggested that receiving feedback with \textit{a low-scored overall evaluation} encouraged participants to revise their writing more than with a high-scored overall evaluation.
However, we did not observe any significant effect of either \textit{AI-reframed positive summary} or \textit{overall feedback evaluation} on participants' quality of revision, though participants perceived their revision quality was increased when receiving \textit{AI-reframed positive summary}.


\subsection{Factors that Influence Receivers' Motivation to Revise (RQ4)}
\label{sec:RQ4}
Based on the thematic analysis of the open-ended comments in the survey, we identified seven factors that potentially influenced participants' motivation to make revisions. 
\rhl{Please note that we analyzed participants' ``claimed motivation,'' as they may not explicitly articulate their reasons for making or not making revisions in an open-ended survey format. 
Additionally, this list of motivational factors is not exhaustive; it only includes those we identified based on our current data collection approach.}
We analyzed the data using the \textit{attribution of motivation theory} to understand how people explain their motivation for making revisions.
This analysis allows us to know what aspects can be taken care of if we want to leverage LLMs to polish critiques to support the delivery of peer feedback. 
Among the seven factors, we found that four of the factors were related to \textit{intrapersonal attribution of motivation}, where people attribute their motivation to revise to themselves. We also found that three factors were related to \textit{interpersonal attribution of motivation}, where people attribute their motivation to revise to external things, including the feedback itself and the reviewers.
Note that the thematic analysis was done with all data without comparing the percentage of each time quantitatively across conditions.

We present \textit{intrapersonal factors} that influenced participants' motivation for revision first.

\subsubsection{\textbf{Intrapersonal factor: Actionability of the feedback}}
Participants' decisions to revise or not could be influenced by how feasible they thought they could address the issue in the revision process. 
They would balance their expected quality and efforts to make revisions. 
For instance, P28 shared, \textit{``Overall, I did not want to throw away too much of the initial cover letter, so I tried to add the feedback I thought was easier to insert into the cover letter.''} (P28, \textit{NoPosFramed-HighScored} condition.) 
P70 also stated, `\textit{`Some of the feedback was too generic, so I didn't implement it. When it was specific, I did it.''} (P70, \textit{NoPosFramed-LowScored} condition.)

\subsubsection{\textbf{Intrapersonal factor: Alignment with personal goals/values}}
Some other participants shared that their motivation for revision was influenced by whether and how much the critique aligned with their personal goals or values for the given writing task.
Their motivation to challenge themselves, reflect their authentic self in the writing, or their own standard about the writing can all influence their decision for revision. 
For example, P02 noted, \textit{``I completely ignored the rudest parts. My writing is fine and not too casual for a cover letter, so I didn't listen to that. I did add some more specific information regarding the social media numbers I hit, but I disagree that a cover letter is the right place for that. Concise is better.'' }(P02, \textit{NoPosFramed-HighScored} condition.) 
P126 also shared that \textit{``It depended on the way in which I worded my letter and still being true to myself. ''}(P126, \textit{PosFramed-LowScored} condition.)

\subsubsection{\textbf{Intrapersonal factor: Emotional response}}
Some participants explicitly mentioned their emotions when deciding what to revise or not. 
When emotion was mentioned in the open-ended comments, almost all of them were negative emotions, such as feeling ``overwhelmed'', ``disrespected'', ``disappointed'', ``messed up'', or ``annoyed''.
This result was consistent with the finding in Section \ref{sec:H1a}, which showed that participants experienced negative emotions significantly more when not receiving AI-reframed positive summaries with the critiques.
As exemplified in P88's comment, \textit{``I didn't like the way because I felt attacked by the person who replied to my letter. I only changed a few things about what I wrote initially because I didn't have enough interest in completing the task as it was asked of me.''} (P88, \textit{NoPosFramed-LowScored} condition.)
P76 also described, \textit{``I had feelings of annoyance since I believe that the person who reviewed my essay does not understand my point of view.''} (P76, \textit{NoPosFramed-LowScored} condition.)

\subsubsection{\textbf{Intrapersonal factor: Motivation to learn and improve}}
Participants' motivation to improve themselves and learn new things tended to make them have a positive attitude while interpreting the critiques they received. 
Those participants explicitly mentioned that critiques helped them improve their writing or enabled them to know which aspects should be improved. 
For example, P40 shared, \textit{``I tried to follow all instructions because I felt I could really learn from them and apply them to my real life. I had to write many cover letters when I was searching for a job, [...]. I feel I can improve my writing skills from now on. Thank you.''} (P40, \textit{PosFramed-HighScored} condition)
Similarly, P95 noted, \textit{``I tried to take almost every feedback because it was important to me to improve my letter.''} (P95, \textit{NoPosFramed-LowScored} condition)

Next, we present \textit{interpersonal factors} that influenced participants' motivation for revision.

\subsubsection{\textbf{Interpersonal factor: Perceived quality of the feedback}}
\label{sec:feedback quality}
The perceived fairness and constructiveness also influenced participants' motivation to make revisions. 
Participants often described the feedback as ``on point'', ``reliable'', ``constructive'', ``honest'', ``specific'', or ``fair'' when they decided to follow the feedback for revision. 
For instance, P07 commented, \textit{``Feedback was very constructive and genuine, which helped me to fix the errors and improve my writing.''} (P07, \textit{NoPosFramed-HighScored} condition).
P53 also noted, \textit{``I took most of the feedback and was grateful for the criticism. I appreciated that the faults were not only pointed out but possible solutions were also provided.''} (P53, \textit{PosFramed-HighScored} condition).
This theme, combined with the quantitative finding in Section \ref{sec:H2a}, suggested that high-quality feedback could motivate people to make revisions when receiving critiques.

\begin{table*}[t]
\caption{Distribution of the percentage of each factor that influenced authors' decisions to make revisions.}
\label{tab:motivations}
\resizebox{0.8\textwidth}{!}{%
\begin{tabular}{|llll|rrrr|r|}
\hline
\multicolumn{4}{|l|}{} &
  \multicolumn{4}{c|}{Condition} &
  \multicolumn{1}{l|}{} \\ \cline{5-9} 
\multicolumn{4}{|l|}{\multirow{-2}{*}{Factors that influence feedback revision}} &
  \multicolumn{1}{c|}{\begin{tabular}[c]{@{}c@{}}NoPosFramed-\\ HighScored\end{tabular}} &
  \multicolumn{1}{c|}{\begin{tabular}[c]{@{}c@{}}PosFramed-\\ HighScored\end{tabular}} &
  \multicolumn{1}{c|}{\begin{tabular}[c]{@{}c@{}}NoPosFramed-\\ LowScored\end{tabular}} &
  \multicolumn{1}{c|}{\begin{tabular}[c]{@{}c@{}}PosFramed-\\ LowScored\end{tabular}} &
  \multicolumn{1}{c|}{Total} \\ \hline
\multicolumn{1}{|l|}{{\color[HTML]{663234} }} &
  \multicolumn{2}{l|}{{\color[HTML]{663234} }} &
  Count &
  \multicolumn{1}{r|}{1} &
  \multicolumn{1}{r|}{1} &
  \multicolumn{1}{r|}{4} &
  2 &
  8 \\ \cline{4-9} 
\multicolumn{1}{|l|}{{\color[HTML]{663234} }} &
  \multicolumn{2}{l|}{\multirow{-2}{*}{{\color[HTML]{663234} Actionability of the feedback}}} &
  \% within column &
  \multicolumn{1}{r|}{3.2 \%} &
  \multicolumn{1}{r|}{2.7 \%} &
  \multicolumn{1}{r|}{12.5 \%} &
  5.4 \% &
  5.8 \% \\ \cline{2-9} 
\multicolumn{1}{|l|}{{\color[HTML]{663234} }} &
  \multicolumn{2}{l|}{{\color[HTML]{663234} }} &
  Count &
  \multicolumn{1}{r|}{8} &
  \multicolumn{1}{r|}{2} &
  \multicolumn{1}{r|}{2} &
  4 &
  16 \\ \cline{4-9} 
\multicolumn{1}{|l|}{{\color[HTML]{663234} }} &
  \multicolumn{2}{l|}{\multirow{-2}{*}{{\color[HTML]{663234} \begin{tabular}[c]{@{}l@{}}Alignment with personal \\ goals/values\end{tabular}}}} &
  \% within column &
  \multicolumn{1}{r|}{25.8 \%} &
  \multicolumn{1}{r|}{5.4 \%} &
  \multicolumn{1}{r|}{6.3 \%} &
  10.8 \% &
  11.7 \% \\ \cline{2-9} 
\multicolumn{1}{|l|}{{\color[HTML]{663234} }} &
  \multicolumn{2}{l|}{{\color[HTML]{663234} }} &
  Count &
  \multicolumn{1}{r|}{4} &
  \multicolumn{1}{r|}{1} &
  \multicolumn{1}{r|}{5} &
  3 &
  13 \\ \cline{4-9} 
\multicolumn{1}{|l|}{{\color[HTML]{663234} }} &
  \multicolumn{2}{l|}{\multirow{-2}{*}{{\color[HTML]{663234} Emotional response}}} &
  \% within column &
  \multicolumn{1}{r|}{12.9 \%} &
  \multicolumn{1}{r|}{2.7 \%} &
  \multicolumn{1}{r|}{15.6 \%} &
  8.1 \% &
  9.5 \% \\ \cline{2-9} 
\multicolumn{1}{|l|}{{\color[HTML]{663234} }} &
  \multicolumn{2}{l|}{{\color[HTML]{663234} }} &
  Count &
  \multicolumn{1}{r|}{7} &
  \multicolumn{1}{r|}{3} &
  \multicolumn{1}{r|}{5} &
  2 &
  17 \\ \cline{4-9} 
\multicolumn{1}{|l|}{\multirow{-8}{*}{{\color[HTML]{663234} \textbf{\begin{tabular}[c]{@{}l@{}}Intrapersonal \\ factor\end{tabular}}}}} &
  \multicolumn{2}{l|}{\multirow{-2}{*}{{\color[HTML]{663234} \begin{tabular}[c]{@{}l@{}}Motivation to learn \\ and improve\end{tabular}}}} &
  \% within column &
  \multicolumn{1}{r|}{22.6 \%} &
  \multicolumn{1}{r|}{8.1 \%} &
  \multicolumn{1}{r|}{15.6 \%} &
  5.4 \% &
  12.4 \% \\ \hline
\multicolumn{1}{|l|}{{\color[HTML]{1C4587} }} &
  \multicolumn{2}{l|}{{\color[HTML]{1C4587} }} &
  Count &
  \multicolumn{1}{r|}{3} &
  \multicolumn{1}{r|}{6} &
  \multicolumn{1}{r|}{2} &
  3 &
  14 \\ \cline{4-9} 
\multicolumn{1}{|l|}{{\color[HTML]{1C4587} }} &
  \multicolumn{2}{l|}{\multirow{-2}{*}{{\color[HTML]{1C4587} \begin{tabular}[c]{@{}l@{}}Perceived expertise/effort \\ of the reviewers\end{tabular}}}} &
  \% within column &
  \multicolumn{1}{r|}{9.7 \%} &
  \multicolumn{1}{r|}{16.2 \%} &
  \multicolumn{1}{r|}{6.3 \%} &
  8.1 \% &
  10.2 \% \\ \cline{2-9} 
\multicolumn{1}{|l|}{{\color[HTML]{1C4587} }} &
  \multicolumn{2}{l|}{{\color[HTML]{1C4587} }} &
  Count &
  \multicolumn{1}{r|}{3} &
  \multicolumn{1}{r|}{16} &
  \multicolumn{1}{r|}{5} &
  14 &
  38 \\ \cline{4-9} 
\multicolumn{1}{|l|}{{\color[HTML]{1C4587} }} &
  \multicolumn{2}{l|}{\multirow{-2}{*}{{\color[HTML]{1C4587} \begin{tabular}[c]{@{}l@{}}Perceived quality of the \\ feedback (fairness, constructive)\end{tabular}}}} &
  \% within column &
  \multicolumn{1}{r|}{9.7 \%} &
  \multicolumn{1}{r|}{43.2 \%} &
  \multicolumn{1}{r|}{15.6 \%} &
  37.8 \% &
  27.7 \% \\ \cline{2-9} 
\multicolumn{1}{|l|}{{\color[HTML]{1C4587} }} &
  \multicolumn{2}{l|}{{\color[HTML]{1C4587} }} &
  Count &
  \multicolumn{1}{r|}{0} &
  \multicolumn{1}{r|}{1} &
  \multicolumn{1}{r|}{2} &
  2 &
  5 \\ \cline{4-9} 
\multicolumn{1}{|l|}{\multirow{-6}{*}{{\color[HTML]{1C4587} \textbf{\begin{tabular}[c]{@{}l@{}}Interpersonal \\ factor\end{tabular}}}}} &
  \multicolumn{2}{l|}{\multirow{-2}{*}{{\color[HTML]{1C4587} Tone of the feedback}}} &
  \% within column &
  \multicolumn{1}{r|}{0.0 \%} &
  \multicolumn{1}{r|}{2.7 \%} &
  \multicolumn{1}{r|}{6.3 \%} &
  5.4 \% &
  3.7 \% \\ \hline
\multicolumn{1}{|l|}{} &
  \multicolumn{2}{l|}{} &
  Count &
  \multicolumn{1}{r|}{5} &
  \multicolumn{1}{r|}{7} &
  \multicolumn{1}{r|}{7} &
  7 &
  26 \\ \cline{4-9} 
\multicolumn{1}{|l|}{} &
  \multicolumn{2}{l|}{\multirow{-2}{*}{Others}} &
  \% within column &
  \multicolumn{1}{r|}{16.1 \%} &
  \multicolumn{1}{r|}{18.9 \%} &
  \multicolumn{1}{r|}{21.9 \%} &
  18.9 \% &
  19.0 \% \\ \cline{2-9} 
\multicolumn{1}{|l|}{} &
  \multicolumn{2}{l|}{} &
  Count &
  \multicolumn{1}{r|}{31} &
  \multicolumn{1}{r|}{37} &
  \multicolumn{1}{r|}{32} &
  37 &
  137 \\ \cline{4-9} 
\multicolumn{1}{|l|}{\multirow{-4}{*}{}} &
  \multicolumn{2}{l|}{\multirow{-2}{*}{Total}} &
  \% within column &
  \multicolumn{1}{r|}{100.000 \%} &
  \multicolumn{1}{r|}{100.000 \%} &
  \multicolumn{1}{r|}{100.000 \%} &
  100.000 \% &
  100.000 \% \\ \hline
\end{tabular}%
}
\end{table*}

\subsubsection{\textbf{Interpersonal factor: Perceived expertise/effort of the reviewers}}
\label{sec:reviewer's expertise}
Some participants mentioned that their motivation to revise was mainly due to the perceived expertise or the perceived effort of the reviewers. 
For those who mentioned the perceived characteristics of the reviewers, most of them thought reviewers were ``skilled'', ``trustworthy'', ``professional'', ``experienced'', ``knowledgeable'', or ``making excellent points''.
As P123 shared, \textit{\rhl{``I was very happy with the feedback as it highlighted all the aspects that my writing lacked. I realized that the reviewer was very skilled in the points and clear explanations they provided. Adding to what I previously lacked made me feel that the cover letter now stood a greater chance of being approved.''}} (P123, \textit{PosFramed-LowScored} condition)
This theme, combined with the quantitative finding in Section \ref{sec:H2b}, suggested that the perceived expertise of the reviewers could also play a part in motivating people to make revisions when receiving critiques.

\subsubsection{\textbf{Interpersonal factor: Tone of the feedback}}
In addition to the content of feedback, the tone of the feedback can also influence participants' motivation to make revisions.
When describing the tone of the feedback, participants mentioned that the feedback sounded ``aggressive'', ``lacked empathy'', ``harsh'', or ``authoritarian''.
AS P94 wrote, \textit{``It was quite aggressive and made me less interested in the role [of the job post]. Some positive feedback is much nicer, and the delivery was very poor.''} (P94, \textit{NoPosFramed-LowScored} condition)
Moreover, P96 described, \textit{``The feedbacks were on point but lacked a bit of empathy. I didn't lie in my cover letter regarding my experience, which is not related to this field, and that makes the exercise more difficult, I believe. [...] ''} (P96, \textit{NoPosFramed-LowScored} condition). 

\begin{figure}[t]
  \centering
  \includegraphics[width=\linewidth]{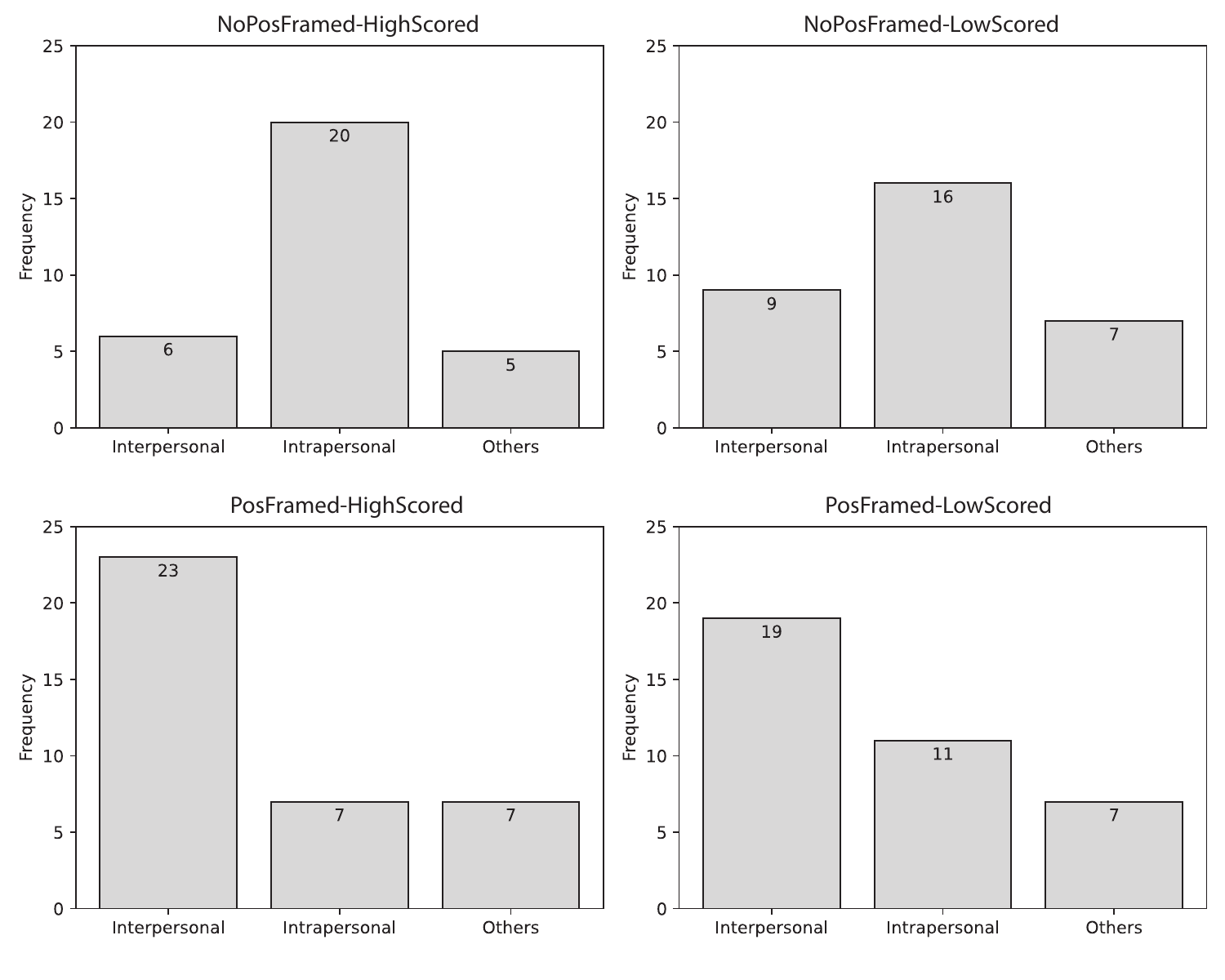}
  \caption{Distribution of different ways participants attribute their motivation for making revisions.}
  \Description{TEMP.}
  \label{fig:motivation}
\end{figure}

Note that all factors we presented above were found in all conditions, though the distribution of the factors was slightly different across conditions.
\autoref{tab:motivations} and \autoref{fig:motivation} showed the distribution of each factor under each condition.
In \autoref{tab:motivations}, we quantified each theme for each condition to show its distribution so readers could have extra information to interpret the themes we identified above.  
In \autoref{fig:motivation}, we merged ``Actionability of the feedback'', ``Alignment with personal goals/values'', ``Emotional response'', and ``Motivation to learn and improve'' together to label them as ``Intrapersonal factor''.
For ``Interpersonal factor'', we merged ``Perceived expertise/effort of the reviewers'', ``Perceived quality of the feedback (fairness, constructive)'', and ``Tone of the feedback'' together.
Although we could not come to any conclusion about whether an AI-reframed positive summary or showing an overall evaluation significantly influenced certain ways for participants to make attribution of motivation for making revisions, this qualitative analysis gave us insights into what factors influence authors' motivation to iterate their writing when receiving critiques. 
In the next section, we discuss ways to leverage LLMs to polish or reframe critiques based on seven common factors that influence people's attribution of their motivation for making revisions.



\section{Discussion}
Our results showed that receiving critique with an AI-reframed positive summary and seeing a high overall evaluation led authors to have positive emotions.
The AI-reframed positive summary also enabled authors to experience increased autonomy and competence, and they felt that both the quality of the feedback itself and the fairness of the reviewers were better than only receiving critical feedback.
Surprisingly, the AI-reframed positive summary did not have much influence on the revision outcome.
Instead, receiving a low overall evaluation made authors revise more than receiving a high overall evaluation.
Through the thematic analysis, we also identified seven key factors that influenced authors' motivation to make revisions or not after receiving critiques.
We summarized our key findings in \autoref{tab:summary}.

\subsection{The Effect of Showing AI-reframed Positive Summary on Critique Reception at Perceptual-Level}
Consistent with prior literature documenting that the use of positive language can encourage authors when receiving critiques \cite{Nguyen2017fruitfulfeedback}, our study extends these findings by showing that reframing critiques by highlighting the positive aspects further enhances critique reception (Section \ref{sec:RQ1}, \ref{sec:RQ2}).
Past research has shown that incorporating praise alongside critiques can help soften the tone of criticism  \cite{hyland2001sugaringthepill_2001}, reinforcing appropriate behaviors, reducing the perceived threat in feedback giver-receiver relationships, and fostering the feedback receivers’ self-esteem \cite{cavanaugh2013performanceFeedback_2013}.
However, related works also suggest that general praise or veiled praise that shows indirect criticism makes authors feel confused and discourages revision \cite{cardelle1981WrittenFeedback}. 
Instead of directly praising the authors, our proposed AI-reframed positive summary was designed to help people see the positive side of criticism while keeping reviewers' original critiques.
The result suggested that positive summaries reframed by AI not only softened the critiques but also empowered authors in the revision process through increased autonomy and competence.

Interestingly, appending an AI-reframed positive summary to critiques also positively influenced authors' perception of the critique and reviewers (Section \ref{sec:RQ2}, Section \ref{sec:feedback quality}, and Section \ref{sec:reviewer's expertise}).
This finding highlights once again that peer feedback not only helps individual authors improve their work but also establishes a positive impression of the review itself and reviewers.
This positive perception could lay a strong foundation for establishing trust with other peer reviewers/experts, which can benefit the research community or the peer feedback platforms more broadly in the long run.
Both anecdotal evidence and the literature have suggested that low-quality and poor peer review practice caused a negative impact on the evolution of science, including demotivating researchers from staying in a research community \cite{McCloskey2022PilotPoorReview, silbiger2019unprofessional, horta2024crisis}.
Thus, to prevent people from getting discouraged by critical feedback during peer reviews and potentially leaving the research community, which could result in only those who are emotionally tough or indifferent prevailing, we invite the research community to think about the possibility of integrating AI-generated positive summaries with the critical feedback in peer reviews. 
This would help to preserve diversity in the research community and broaden the way topics are studied \cite{schneider2001search}.

\rhl{Meanwhile, we should make AI's involvement transparent to the authors when deploying AI-assisted feedback outside the controlled experiment setting.
Literature on AI-mediated communication has shown that distrust from receivers to senders arises when receivers are aware that AI was involved in generating text that could also be generated by a mixture of humans and AI \cite{jakesch2019aimc}. 
To alleviate potential distrust that authors might feel towards reviewers when integrating AI-reframed positive summaries into the peer review process, one possible strategy is to present the AI-generated section separately from the human-written review, while clearly indicating AI's role. 
Although the impact of revealing AI's involvement in AI-assisted peer feedback has not been explored in the current scope, and we could only disclose the use of AI at the end, future research can examine the effects of disclosing AI's role at different review writing stages in the peer review process. 
This could lead to a better understanding of how such disclosures affect the receivers' (authors') trust toward the sender (reviewer) in AI-assisted peer feedback.
}

Unexpectedly, receiving a positive summary reframed by AI only changed authors' perceptions, not their actual quality of revision (Section \ref{sec:RQ3}).
Prior research also did not find any significant improvement in the quality of writing when authors received either positive (i.e., praise) or negative (i.e., blame) feedback, but positive feedback led to authors' increased favorable attitudes and motivation \cite{taylor1966PraiseCreativeWriting}.
One possible reason for the lack of improvement in revision quality in our study may be that the cognitive reframing strategy we employed was quite generic. 
In the current design of the prompt for the AI-reframed positive summary, we encouraged participants to adopt a positive attitude and consider the positive aspects of a negative situation.
However, the effectiveness of cognitive reframing could vary depending on whether individuals' learning styles \cite{van2012learningStyle} match with the specific cognitive reframing strategy or depending on the different focus of reframing strategies, such as rationality or actionability \cite{sharma2023cognitive}.
Based on our results, future work could explore different reframing strategies for LLMs to generate various types of reframed summaries. 
This could involve considering authors' learning styles or preferred thinking styles and examining how different types of AI-reframed positive summaries lead to specific behavioral changes in the context of peer feedback.

\begin{table*}[h!]
\caption{Research questions and summary of the key findings}
\label{tab:summary}
\resizebox{0.8\linewidth}{!}{%
\begin{tabular}{|l|l|}
\hline
\textbf{Research questions} &
  \textbf{Main findings} \\ \hline
\begin{tabular}[c]{@{}l@{}}RQ1: How does the combination of the presence of \\AI-reframed positive summary (present/absent) and \\different overall evaluations (high/low) affect \\the authors' intrapersonal perception when \\receiving critical feedback?\end{tabular} &
  \begin{tabular}[c]{@{}l@{}}\textbf{H1a}: Receiving AI-reframed positive summary \\ and a high overall evaluation of the feedback \\ led people to \textbf{experience a positive emotion}. (supported)\\ \textbf{H1b}: Whether receiving AI-reframed positive summary \\ 
  or receiving high/low overall evaluation \\ did not change people’s \textbf{self-efficacy in writing} when \\ receiving critical feedback. (not supported)\\ \textbf{H1c}: Receiving critical feedback came with an \\ AI-reframed positive summary \textbf{increased people’s} \\ \textbf{autonomy and competence in making revisions}. \\ (supported)\end{tabular} \\ \hline
\begin{tabular}[c]{@{}l@{}}RQ2: How does the combination of the presence of \\AI-reframed positive summary (present/absent) and \\different overall evaluations (high/low) affect \\the authors’ interpersonal perception when \\receiving critical feedback?\end{tabular} &
  \begin{tabular}[c]{@{}l@{}}\textbf{H2a}: Receiving AI-reframed positive summary \\ \textbf{increased the perceived fairness of the critical feedback}. \\ (partially supported)\\ \textbf{H2b}: Receiving AI-reframed positive summary \textbf{enhanced} \\ \textbf{the perceived fairness of the feedback} \\ \textbf{provider}. (partially supported)\end{tabular} \\ \hline
\begin{tabular}[c]{@{}l@{}}RQ3: How does the combination of the presence of \\AI-reframed positive summary (present/absent) and \\different overall evaluations (high/low) influence \\the authors' revision outcomes when \\receiving critical feedback?\end{tabular} &
  \begin{tabular}[c]{@{}l@{}}Receiving feedback with a low-scored overall evaluation\\ encouraged people to \textbf{revise their writing more} than with \\ a high-scored overall evaluation. Although little influence \\ was found on participants’ revision quality, participants \\ \textbf{perceived their revision quality was significantly better} \\ when receiving an AI-reframed positive summary.\end{tabular} \\ \hline
\begin{tabular}[c]{@{}l@{}}RQ4: What factors influence the authors' motivation to \\revise their writing when receiving critical feedback?\end{tabular} &
  \begin{tabular}[c]{@{}l@{}}\textbf{Intrapersonal factors:} Actionability of the feedback, \\ alignment with personal goal/value, \\ emotional response, motivation to learn and improve; \\ \textbf{Interpersonal factors:} perceived quality of the feedback, \\ tone of the feedback, perceived expertise/effort \\ of the reviewer.\end{tabular} \\ \hline
\end{tabular}
}%
\end{table*}

\subsection{The Effect of Showing Low Overall Evaluation on Critique Reception at Behavioral-Level}
Our result showed that regardless of receiving AI-reframed summaries, when people received a low overall evaluation for their writing, they made more revisions than those who received a high overall evaluation (Section \ref{sec:RQ3}).
Prior research has shown that feedback highlighting discrepancies between the authors' current performance and their desired standards can lead to more substantial efforts for authors to close the gap \cite{kluger1996FeedbackIntervention}. 
In our study, participants who received low scores might be able to directly see the gap between ``where they are'' and ``where they aim to be'' \cite{hattie2007power}, thus taking feedback more seriously and engaging more deeply with the revision process. 
Past studies suggested that individuals who receive negative feedback often engage in deeper cognitive processing to understand the topic and rectify their mistakes \cite{hattie2007power}. 
Thus, it is possible that receiving a low-scored overall evaluation made authors scrutinize the feedback more thoroughly and revise extensively to reduce the discrepancies between their writing and expected goals.

Though receiving a low-scored overall evaluation led to authors' increased revisions, receiving too much negative feedback can be harmful and discouraging, especially for those who have low self-efficacy in writing \cite{pajares1994confidence, taylor1966PraiseCreativeWriting}.
It has been found that people with low writing self-efficacy were more likely to be negatively affected by critical feedback \cite{pajares1994confidence}. 
These writers showed decreased motivation and less engagement in revising their work, indicating that their perception of their abilities influenced how they responded to feedback \cite{pajares1994confidence}.
When authors doubt their capabilities, overwhelming criticism can lead to feelings of inadequacy, reduced motivation, and possibly a decline in writing performance.
Therefore, we suggest that when reviewers want to motivate authors to significantly revise their writing during the revision process, combining a low overall score critique with positive reframed summaries could be effective in maintaining the right balance between authors' perceptions and revision behavior.

\subsection{Incorporating Intrapersonal and Interpersonal Factors in AI-assisted Peer Feedback}
Based on the seven motivational factors we identified in Section \ref{sec:RQ4} and the distribution of the motivational factors in \autoref{tab:motivations}, it revealed that authors considered both \textit{intrapersonal} and \textit{interpersonal} factors when deciding whether to make revisions.
Current practice in peer feedback often provides reviewers with a rubric to guide the review writing process.
Our findings pointed out that in addition to following rubrics, incorporating \textit{intrapersonal} and \textit{interpersonal} factors into peer reviews or meta-reviews might be an effective way to motivate authors to make revisions.
For example, future AI-assisted peer review systems could scaffold and help reviewers as they write meta-review by incorporating intrapersonal factors, such as ``enhancing the actionability of the feedback,'' and ``encouraging authors to make improvements,'' as well as interpersonal factors, such as ``reframing the critique in a friendly tone,'' and ``showing the efforts reviewers have put into writing reviews'', into the meta-review.

\rhl{Meanwhile, we acknowledge that we could not fully capture all possible motivational factors with the current online controlled experiment. 
Although we attempted to qualitatively analyze the claimed motivators and present their quantified distribution, future studies could explore motivations by building on the factors identified in this study and incorporating additional relevant measures (e.g., semi-structured interviews) to enhance the robustness of the findings.
}

\section{Limitations and Future Work}
Though we identified the positive impact of AI-reframed positive summaries on critique reception, we acknowledge that our study has some limitations.
Firstly, we designed a writing task that asked participants to write a cover letter to apply for a job post.
In light of the potential challenges of having researchers write about their research and then being processed by GPT, which could compromise the confidentiality of their work during the study, we have chosen to employ a writing task that can be well-controlled in an experimental setting.
Writing a cover letter allowed participants to write something that they care about to a certain extent so that we can simulate the situation where people receive critique for something that matters to them.
Moreover, writing a cover letter is a relatively short task that can be completed within a reasonable amount of time.
However, we acknowledge that writing a cover letter is different from writing research articles, and we need to be cautious when interpreting the generalizability of our findings. 
To enhance the generalizability of our findings, future work could expand the scope of the study from a controlled experiment to a large-scale field study or explore the possibility of analyzing the dataset of peer review from some research communities. \rhl{ With a field study or dataset analysis, we may also have a sufficiently large sample size to include the factor of initial writing quality and enhance the current understanding.}

\rhl{
Secondly, though we found that appending a positively reframed summary to peer feedback supported critique reception, future studies can explore the best way to present the positively reframed content. We appended the positively reframed summary at the end of the critique in the current setting, which may be influenced by recency bias \cite{anderson1963effects, forgas2011recencyeffect}, where people tend to give greater importance to recent events (i.e., positive summary) than historical ones (i.e., critical feedback). Future work can examine whether recency bias occurs in a peer review context and how to cope with it with the suitable presentation of the positively reframed summary.
}

This study explored and evaluated the possibility of using LLMs to reframe human-written critiques in a positive way and demonstrated its positive impact on authors' emotions, perception of the critique, and perception of the reviewer. 
Based on the current finding, we encourage the research community to explore different ways in which LLMs could be beneficial in supporting reviewers in delivering constructive peer feedback together. 
For instance, incorporating an option in the peer review system for reviewers and/or authors to activate the AI-reframed feature while composing and receiving critiques.

\section{Conclusion}
When critiques are delivered in a harsh tone in peer review, it can make people feel negative emotions and discourage them from revising their work.
We proposed using AI to reframe human-written critique in a positive way and investigated its effect on the perception of the critique when it is combined with high or low overall evaluations.
We found that adding AI-reframed positive summaries to critique made authors experience positive emotions, increased autonomy and competence, and perceive both the critique and reviewers as fair.
In contrast, receiving low-scored overall evaluations increased the amount of revisions people made.
Furthermore, people tended to attribute their motivation for making revisions to interpersonal factors when receiving AI-reframed positive summaries.
We discussed the opportunity for combining a low-scored overall evaluation and an AI-reframed positive summary of critical feedback for a constructive and friendly peer review process.






\begin{acks}
We extend our gratitude to all the anonymous participants who took part in this study. 
We also appreciate the constructive feedback provided by the reviewers, which has significantly enhanced this work. 
Lastly, we would like to thank Jiayi Yang and Xiaotong Li for their valuable assistance in evaluating all the written outcomes from the participants.
\end{acks}

\bibliographystyle{ACM-Reference-Format}
\bibliography{0-References}










\end{document}